\documentclass{aastex62}
\newcommand{\ltsim}{\raisebox{-1.0ex}{$\stackrel{\textstyle<}{\sim}$}}
\def\kms{km~s$^{-1}$}
\def\al{Alfv\'{e}n}
\def\goes{{\sl GOES}}

\def\hinode{{\sl Hinode}}

\def\p78{{\sl P78-1}}

\def\stereo{{\sl STEREO}}
\def\sdo{{\sl SDO}}

\def\smei{{\sl SMEI}}

\def\al{Alfv\'{e}n}

\def\kms{km~s$^{-1}$}

\begin{document}
%

\title{Coronal-Jet-Producing Minifilament Eruptions as a Possible Source of Parker Solar Probe (PSP) Switchbacks}

\author{Alphonse C. Sterling}
\affiliation{NASA/Marshall Space Flight Center, Huntsville, AL 35812, USA}

\author{Ronald L. Moore} 
\affiliation{NASA/Marshall Space Flight Center, Huntsville, AL 35812, USA}
\affiliation{Center for Space Plasma and Aeronomic Research, \\
University of Alabama in Huntsville, Huntsville, AL 35805, USA}

\begin{abstract}

The Parker Solar Probe (PSP) has observed copious rapid magnetic field direction changes in
the near-Sun solar wind.  These features have been called ``switchbacks," and their
origin is a mystery.  But their widespread nature suggests that they may be generated by a 
frequently occurring process in the Sun's atmosphere.  We examine the possibility that the switchbacks 
originate from coronal jets.  Recent work suggests that many coronal jets result when photospheric
magnetic flux cancels, and forms a small-scale ``minifilament" flux rope that erupts 
and reconnects with coronal field.  We argue that the reconnected erupting minifilament 
flux rope can manifest as an outward propagating \al ic fluctuation that steepens into 
an increasingly compact disturbance
as it moves through the solar wind.  Using previous observed properties of coronal jets 
that connect to coronagraph-observed white-light jets (a.k.a.\ ``narrow CMEs"), 
along with typical solar wind speed values, we expect the coronal-jet-produced disturbances to
traverse near-perihelion PSP in $\ltsim$25~min, with a velocity of $\sim$400~\kms.  To 
consider further the plausibility of this idea, we show that a previously studied 
series of equatorial latitude coronal jets, originating from the periphery of an active region, 
generate white-light jets in the outer corona (seen in \stereo/COR2 coronagraph images; 
2.5---15 $R_{\odot}$), and into the inner heliosphere (seen in \stereo/Hi1 heliospheric 
imager images; 15---84 $R_{\odot}$).  Thus it is tenable that disturbances
put onto open coronal magnetic field lines by coronal-jet-producing erupting minifilament flux ropes
can propagate out to PSP space and appear as switchbacks.

\end{abstract}


\section{Introduction}
\label{sec-introduction}

The Parker Solar Probe (PSP) mission \citep{bale.et16,fox.et16,kasper.et16} has for the 
first time carried out {\it in situ} observations in the near-Sun solar wind, reaching $\sim$35
$R_\odot$ in 2018 November and also in 2019 April.  An exciting early observation from the
mission is that the near-Sun magnetic field is replete with transient, kinked structures that
have been called ``switchbacks"
\citep{bale.et19,kasper.et19,dudok.et20,mozer.et20}.  Similar structures were also seen
earlier \citep[e.g.,][]{kahler.et96,yamauchi.et04,suess07}.    The source of these features
is not understood.  A possibility that we investigate here is that solar coronal jets might be
responsible for the switchbacks \citep[as suggested by, e.g.,][]{horbury.et20}.  

Here we examine the possibility that a recently suggested process for making coronal jets, based on
the eruption of small-scale filaments and their enveloping field that reconnects with coronal
field,  results in the switchbacks.

\section{Coronal Jets and White-light Jets}
\label{sec-jets}

\subsection{Coronal Jets}
\label{subsec-coronal_jets}

Coronal jets have been observed for some time at X-ray
\citep[e.g.][]{shibata.et92,cirtain.et07} and EUV   \citep[e.g.][]{nistico.et09}
wavelengths.  They are frequently occurring phenomena, with a rate of about 60/day in polar
coronal holes alone \citep{savcheva.et07}.  For reviews of jets,  see \citet{shibata.et11},
\citet{raouafi.et16}, and \citet{hinode.et19}.

Recent observations support that at least many, if not most or all, coronal jets result  from
the eruption of a small-scale filament, or {\it minifilament}, and its enveloping magnetic  field.
\citet{sterling.et15} proposed a ``minifilament-eruption  model" for coronal jets, and argued that the
entire coronal-jet event is a scaled-down  version of the larger-scale eruptions that create typical
solar flares and CMEs.   Apparently almost all  coronal jets, at least those in quiet Sun and coronal
hole regions, are produced by such eruptions.  Often the small-scale erupting field contains cool
material (appearing as the minifilament)  in the core of the erupting magnetic arcade  
\citep[e.g.][]{hong.et14,moore.et10,shen.et12,shen.et17,shen.et19,mcglasson.et19}, where the  eruption
can be either ejective or confined \citep{sterling.et15}.  We cannot however totally rule out that
some other process, such as the much-earlier-suggested emerging-flux mechanism 
\citep{shibata.et92,yokoyama.et95}, might produce some jets and expel cool material into the
corona.

Other observations show that the coronal jets originate at photospheric locations where magnetic flux 
cancelation occurs under the pre-eruption minifilament
\citep[e.g.][]{shen.et12,hong.et14,young.et14a,young.et14b,panesar.et16a}.  We have  found
observational evidence  that in many cases magnetic flux cancelation creates the minifilament flux
rope and triggers the eruption of the flux rope and its enveloping magnetic arcade, and this eruption
produces the coronal jet 
\citep{panesar.et16a,panesar.et17,sterling.et17,panesar.et18a,mcglasson.et19}. An alternative view
argued by \citet{kumar.et19} is that often shearing and/or rotational photospheric motion is
responsible for the build up of energy along the minifilament channel that gets released through
eruption and produces the jet.

Figure~1 shows the basic minifilament-eruption jet-production idea of \citet{sterling.et15}.  
Figure~1(a) shows a 
cross-sectional view of a 3D positive-polarity anemone-type field inside of a majority 
negative-polarity ambient open
field.  One side of the anemone is highly sheared (and often twisted)
and contains a minifilament (blue circle).  In Figure~1(b) the minifilament field is erupting and
undergoing reconnection in two locations: (1) {\it internal} (``tether-cutting'' type)
reconnection (larger red X), with the solid red lines showing the resulting reconnected
fields, and where the thick red semicircle represents the ``jet bright point'' (JBP) at the jet's base;
and (2) {\it external} (a.k.a.\ ``interchange'' or ``breakout'') reconnection occurs at the site of the
smaller red X, with the dashed  lines indicating its two reconnection products.  Figure~1(c) 
shows that if the
external reconnection proceeds far enough, then the minifilament material can leak out onto
the open field.  Shaded areas represent heated jet material visible in X-rays and
some \sdo/AIA EUV channels as the jet's spire. This picture has been successfully simulated by
\citet{wyper.et17,wyper.et18a} (they refer to this ``minifilament-eruption model for jets"
as a ``breakout model for jets," since breakout-type reconnection is integral to the jet's
production).

Active region (AR) coronal jets similarly show evidence that they are made from small-scale eruptions,
and that these eruptions are prepared and triggered by magnetic flux cancelation.  It seems 
however as if the eruptions leading to AR jets less frequently (than in non-AR areas) 
carry cool material that appears as a minifilament, although evidence 
indicates that a minifilament-type flux-rope field still erupts to make the AR jets
\citep{sterling.et16b,sterling.et17}.

\subsection{White-light Jets and Twists on Coronal Jets}
\label{subsec-twists_on_cmes}

Coronal jets are capable of producing features observed in coronagraphs called ``narrow
CMEs" or ``white-light jets''
\citep[e.g.,][]{wang.et98,nistico.et09,moore.et15,sterling.et16b}. Such studies
showed a clear connection between coronal jets on the Sun and the white-light jets observed
with either the \stereo\ COR1 coronagraph \citep{nistico.et09,nistico.et10,paraschiv.et10},
or in the LASCO C2 coronagraph \citep{wang.et98,moore.et15,sterling.et16b}.  In other
cases, jets can apparently propel outward -- or at least accompany -- broader ``bubble-like"
CMEs
\citep[e.g.][]{bemporad.et05,shen.et12,alzate.et16,panesar.et16b,miao.et18,duan.et19,solanki.et19};
our focus here however is on the narrow CMEs.

Several studies have found twist on jets \citep[e.g.][]{pike.et98}.  A few such investigations
have  measured the number of turns a jet undergoes over its lifetime; \citet{shen.et11a} found a jet
to undergo 1.2 to 2.6 turns, while \citet{chen.et12} estimated the same jet to undergo 3.6 turns. 
\citet{hong.et13} estimated a different jet, one
that may have produced a white-light jet, to undergo 0.9 turns. \citet{moore.et13} found that 24 of 29
(83\%) random polar jets that they examined had one-half or fewer turns, while the remaining five events
had up to 2.5 turns.   \citet{liu.et19} studied 30 off-limb ``large-scale rotational" coronal jets,
and found that they all underwent at least 1.3 turns and 80\% of them rotated less than 2.8 turns,
with the one with maximal rotation having 4.7 turns.  References in \citet{liu.et19} discuss other
papers with jet-twist measurements.

\citet{moore.et15} studied 14 jets that produced white-light jets, and found that they  had twist
values of one-half to 2.5 turns.  They argued that an erupting twisted flux rope (which in subsequent
papers we argue is a minifilament flux rope) can inject twist onto the white-light jet.  A conclusion
of their study  was that all of the coronal jets that made white-light jets in their study had
comparatively large amount of twist in the spire of the coronal jets when observed in AIA 304~\AA\@. 
Thus it was apparent that the twist was an important factor for the coronal jets to make it out to a
few $R_\odot$ into the corona.  


Figure~2 shows our picture for how a coronal-jet-producing minifilament eruption could launch a
white-light jet.  Initially the minifilament field that erupts to form the coronal jet would carry twist,
as in Figure~2(a).  When this twisted erupting flux rope strikes ambient
field of  opposite polarity in Figure~2(b) (corresponding to Fig.~1(b)) and 
undergoes external reconnection, that reconnection transfers twist onto the ambient open
field, as proposed by  \citet{shibata.et86}.  This twist would propagate outward (Fig.~2(c)) 
as an \al ic twist-wave packet, driving the white-light jet  seen in coronagraph images. 
Eventually (2(d)) the near-original setup is recovered, but with the imparted twist from 
the reconnected minifilament field now removed from in and near the jet's base field.

\section{Possible Production of Switchbacks by Propagating Magnetic Twist on White-light Jets}
\label{sec-switchbacks}

Figure~3 is a continuation of Figure~2, showing how the twist imparted to an open field
by a coronal-jet-producing minifilament eruption evolves into a switchback, where the 
yellow circles represent the Sun, and the blue lines represent heliospheric field lines that
are  curved, following a Parker spiral, with respect to a radial line (black).  The twist put 
onto the white-light jet (Fig.~2(c)) will continue to propagate out into the heliosphere.  
In Figure~3(a), the twist is shown as an extension to the situation in Figure~2(c), with the 
twist having about the same small pitch angle as seen in the C2 images of Fig.~6 of 
\citet{moore.et15}.

In Figure~3(a), the twist imparted to the ambient coronal field in Figure~2(c)
continues  to propagate outward, becoming the red disturbance that appears as a low-pitch
twist wave packet moving outward (the radial extent of the twist packet would be comparable 
to a solar radius, and so is exaggerated by a factor of $\sim$5 compared to the Sun  in this
schematic representation).  Figure~3(b) shows  the pitch of the disturbance
increasing as it moves further from the Sun.  This is our expectation, because it moves into
a regime with progressively lower \al\ velocity. (In the corona,  the \al\ velocity, 
$V_A \approx 1000$~\kms.  At the first PSP perihelion, \citeauthor{bale.et19}~\citeyear{bale.et19} report
$V_A \sim$100~\kms\ in the solar wind at 36.6 $R_\odot$.) 
Based on \citet{moore.et15}, the disturbance in C2 has length $L$ comparable to $R_\odot$.
The front of the disturbance moves more slowly than its rear, resulting in a
``compression" (increasing  pitch angle) of the disturbance.  In Figure~3(c), this
pitch-angle steepening of the disturbance continues  as it moves even further from the Sun,
appearing as a ``switchback" by the time it encounters PSP\@.

PSP would detect the \al-wave packet as the packet flows and propagates by.  The radial speed of the
packet will vary depending on its distance from the Sun.  At the time of its launch in the low corona,
the packet would have a speed of about that of the local coronal \al\ speed ($\sim$1000~\kms),  with a
solar wind velocity, $V_{SW}$, of practically zero.  At PSP, the \al\ velocity will be $\sim$100~\kms\
as mentioned above, but it will  be riding in the solar wind with $V_{SW}\approx 300$~\kms (which is
the baseline solar-wind speed reported by \citeauthor{kasper.et19}~\citeyear{kasper.et19} during the
first PSP perihelion passage); that is, it will pass PSP at about 400~\kms.

The length of the packet, $L$, at the Sun will be about $V_A \times \tau$, where we can take 
$\tau \approx 600$~s, since a typical coronal jet lasts about ten minutes
\citep[e.g.,][]{savcheva.et07}. So the pulse's length near the Sun, $L_{cor}$, would be 
$L_{cor}\sim$600{,}000~km.  At PSP, a packet
of this length traveling at 400~\kms\  would appear as a pulse passing the spacecraft in 
1500~s, i.e.\ $\sim$25~min. 
The \al-wave-packet's length at the spacecraft, $L_{PSP}$, however  will be reduced from
what it was in the corona, via the above-argued pitch-angle-steepening rationale. Thus the
passage of the pulse (the switchback) past PSP should be less than about 25~min.
Smaller-scale ``network jets" (or ``jetlets") \citep[e.g.][]{raouafi.et14} appear to work
like typical coronal jets \citep{panesar.et18b}.  Thus these smaller events plausibly 
produce many briefer switchbacks in the solar wind.  Observed switchbacks have durations
ranging from less than a second to more than an hour \citep[e.g.][]{dudok.et20}.

\section{Observations of Coronal Jets in the \stereo\ Outer Corona}
\label{sec-stereo}

While \S\S\ref{sec-jets} and~\ref{sec-switchbacks}  present a scenario whereby
coronal jets might theoretically make PSP switchbacks, there still remains the question of
whether coronal-jet effects can actually propagate out to the distances of tens of solar radii where
they might be detected by PSP\@.  As pointed out in \S\ref{sec-introduction}, there have
been several observations of  the effects of coronal jets out to the \stereo/COR1 (1.5---4
$R_{\odot}$;  \citeauthor{howard.et08}~\citeyear{howard.et08}) and LASCO~C2 (1.5---6
$R_{\odot}$) distances.  Polar coronal  jets have been tracked even further, into the
\stereo/COR2 (2.5---15 $R_{\odot}$) field of view (FOV), and then as  density enhancements at
substantial fractions of an A.U.\ in 3D reconstructions from Solar Mass Ejection Imager 
(\smei) data in recent studies \citep{yu.et14,yu.et16}.  

In this section we present observations of another example of the signatures of coronal jets 
propagating into the
outer corona and inner heliosphere.  Our example differs from those of
\cite{yu.et14,yu.et16} and \citet{moore.et15}, in that their examples originate from polar
coronal hole jets, while our examples here originate from coronal jets at equatorial latitudes and from the
periphery of an active region.  Our  coronal jets are the same as those of \citet{sterling.et16b},
and that paper showed the jets connecting to white-light jets in the \stereo/COR1 FOV\@.  Here we show that some
of the coronal-jet signatures can be tracked to locations farther from the Sun.

\subsection{Coronal-Jet Origins}
\label{subsec-origins}

We give a brief summary of the solar origins of the coronal jets, more details of which are
provided in \citet{sterling.et16b}. That paper studied a series of coronal jets that occurred at the
edge of NOAA AR~11513.  While they primarily used \sdo/AIA  data for their analysis, they
also used complementary views from \stereo-B and showed that many of their coronal jets produced white-light
jets in the \stereo-B COR1 FOV\@.  While the AIA images showed that the COR1 features originated
from several locations around the AR, here we concentrate on the features that made
white-light jets in COR1 at a position angle of $\approx$315$^\circ$; this is because it is
at about this same position angle where we can identify white-light jets further out in the corona.
From the COR1 coronagraph video of \citet{sterling.et16b} (the video accompanying Fig.~5 in
that paper), it can be seen that the white-light jets from this position angle largely originated from location
of the AR labeled ``C" in that paper \citep[see Fig.~3(a) of][]{sterling.et16b}. Hence we
concentrate on coronal jets from that location in the following.

Figures~4(a---c), and accompanying video vid4abc, show coronal jets from this location in
AIA~304~\AA, and  Figures~4(d---f) and accompanying video vid4def show the magnetic
evolution of the region in  \sdo/HMI magnetograms.  Table~1 lists the primary jets occurring
from this location over 19:00---23:50~UT on 2012 June 30, which is the time period we will
focus on. Figures~4(a---c) track the progress of jet J5 of table~1.

As discussed in \citet{sterling.et16b}, the coronal jets from this location originate from either of the
two neutral lines pointed to by the yellow and red arrows in Figure~4(d).  Over the time of
Figures~4(d---f), the positive-polarity patch between these arrows decreases in size; from video
vid4def, this decrease is consistent with convergence of the positive-polarity flux patch and 
surrounding negative-polarity flux, resulting in flux cancelation.  From this observation, in
conjunction with our understanding of coronal-jet initiation outlined in \S\ref{sec-introduction},  we
conclude that it is likely that flux cancelation built a  minifilament field that erupted to make
the coronal jets, following the picture of Figure~1.  The continued cancelation is responsible for the
continuing series of essentially homologous coronal jets \citep{panesar.et16b,sterling.et17}.

In their study of 14 polar coronal hole jets, \citet{moore.et15} found that coronal jets that 
extended into white-light jets in the LASCO/C2 FOV tended  to be those with relatively high amounts of twist when observed in
AIA 304~\AA\ images.  Those jets reaching C2  had between 0.5 and 2.5 axial  turns, with a peak
near 1.5 turns.  In contrast, they found that a more general population  of 29 jets had axial
rotations mostly between zero and 0.5 turns.  Thus the jets that reach  C2 preferentially have
more twist than the general population of coronal hole jets.  

Our coronal jets here are from an AR rather than a coronal hole, but we can ask whether these jets
show spinning motions.  Inspection of the 304~\AA\ movie vid4abc shows that several jets indeed appear
to spin during their onset phase.  We estimate the number of turns that each jet makes using same
basic procedure  as in \citet{moore.et15}, specifically by picking a feature on the jet, tracking its
lateral motion, and counting how many apparent oscillations it  makes in the left-right (east-west)
direction it makes during the early part of the jet. The black arrows in Figure~4(a---c) show
an  example, where we track an absorbing feature in jet J5 of table~1.  Table~1 provides our results,
giving our estimated number of turns for each jet. Other than jet J3, all of the jets show obvious
indications of spin, where the values range from 0.25 to 1.5 turns, with an average of 0.8 turns. 
Only two of the eight jets (J3 and J6) have spin values smaller than the 0.5 lowest value of the
\citet{moore.et15} coronal jets that made white-light jets.  

Even though our interpretation of coronal-jet spin is based on visual inspection only, there is
strong evidence from spectral studies providing evidence from Doppler measurements that many jets
truly spin \citep[e.g.][]{pike.et98,kamio.et10}.  Similar to the situation in \citet{moore.et15},
the appearance is that the spinning is an {\it unwinding} of the field containing the cool
304~\AA\ jet material, as the spinning eventually slows and stops in all of the cases. 

We measured the outflow velocities of the coronal jets over the FOV of Figure~4, by tracking portions
of the jet spire in emission in 304~\AA; the absorbing material (likely erupting-minifilament
material) sometimes moves out at a slower velocity.  Jet J6 has a velocity higher than the others;
this is probably due to a stronger energy release, as it corresponds to an explosive flare of \goes\
level C1.6. \citet{sterling.et16b} also found that this coronal jet extended to a COR1 white-light jet
that was the fastest of their set: 841~\kms.  This is consistent with the study  of
\citet{shen.et11b}, which provides observational evidence that the \goes\ class of a flare is directly
related to the kinetic energy of the accompanying erupting filament.


\subsection{The Jets in the Outer Corona}
\label{subsec-outer_corona}

Figure~5(a) shows the progression of the jets J4 and J5 into the \stereo/COR1 coronagraph
FOV, based on the results of \citet{sterling.et16b} (see Fig.~5 and accompanying video of
that paper; in that paper, our jets J4 and J5 are respectively jets 6 and 7).  Figure~5(b)
shows the jets in the  \stereo-B COR2 coronagraph, and the accompanying video, vid5abcd,
shows that this feature is clearly a continuation of the jet~J4/J5 feature of Figure~5(a). 
From the 5-min-cadence COR1 movie in \citet{sterling.et16b}, these two jets  occur very
closely together  in time in COR1, and so we cannot differentiate between them in the
15-min-cadence COR2 movie (in vid5abcd, the cadence of both the COR1 and COR2 movies
are set to match the cadence of the COR2 movies). Figure~5(c) shows jet~J6 of Table~1 (this
identification between J6 and the COR1 jet was made in 
\citeauthor{sterling.et16b}~\citeyear{sterling.et16b} using the
5-min-cadence COR1 movie).  Figure~5d shows a white-light jet in COR2 from the same time and
position angle; this is either a continuation of jet J4/J5, or it could be jet J6, or a
combination of jets, but the time cadence of COR2 is not high enough for us to determine
which of these is the case.  In the COR2 video, the white-light jet of Figures~5(b) and~5(d) has velocity of
about 800~\kms.

\subsection{The Jets in the Inner Heliosphere}
\label{subsec-hi1_jets}

Figure~5 shows a COR2 image in 5(e), concurrent with a \stereo-B Hi1 image in
5(f).  Hi1 observes the inner  heliosphere with a wide FOV (15---84
$R_{\odot}$; \citeauthor{howard.et08}~\citeyear{howard.et08}), but offset from Sun center; in
Figure~5(f) the Sun is located off of the left side of the panel.  From the accompanying video,
vid5ef, the jet in Figure~5(f) is a continuation of one of, or a combination of some of,
the Table~1 jets that have already left the FOV of Figure~5(e).  We can confirm that the 
location of the jet with the arrow in
Figure~5(f) corresponds to the position angle of the Table~1 jets by using the large-scale
eruption that appears in COR2 at 12:09~UT in vid5ef.  That eruption is very large, and
expands out into a CME that is visible in the Hi1 video from 15:29~UT\@.  This feature is an
unmistakable continuation of the COR2 eruption. In the Hi1 movie, it has a position angle 
slightly smaller than (just clockwise of) that of the Table~1 jets, and this gives us 
confidence that the jet seen in Hi1 at a slightly
larger position angle in Figure~5(f) indeed corresponds to the jets of Table~1.  (The large
eruption beginning at 12:09~UT in COR2 originates from a neutral line to the east of the
images in Fig.~4; in \citeauthor{sterling.et16b}~\citeyear{sterling.et16b}, the source
location is between locations marked  ``A" and ``B" in Fig.~3(a) of that paper.) That
eruption was of a larger scale than those that make the jets at location displayed in
Figure~4.  In the Hi1 FOV, the white-light jet of Figure~5(f) has velocity of
about 750~\kms; to within the uncertainties of our estimate, this can be regarded as 
about the same as the velocity of the white-light jet (or combination of jets) 
observed in COR2 (\S\ref{subsec-outer_corona}).

\section{Discussion}
\label{subsec-discussion}

Because coronal jets are frequent, and because recent work suggests that they are formed when magnetic
flux ropes erupt away from the solar surface and reconnect with coronal field (Fig.~1),  it is natural
to ask whether the coronal jets could be the source of the magnetic ``switchback"  fluctuations
observed by PSP in the near-Sun solar wind.  We have presented a picture (Figs.~2 and~3) by which the
\al ic fluctuations resulting from the magnetic eruptions that produce the jets might  evolve into
switchbacks. We have also presented evidence that jets at equatorial latitudes can reach the outer
corona and the inner heliosphere (Fig.~5), supplementing earlier studies of white-light jets and
solar-wind disturbances from coronal jets from polar regions (\S\ref{sec-stereo}).  Moreover, 
numerical simulations support that  disturbances put onto open field in the corona can persist out to
many solar radii  \citep{tenerani.et20}.

Our idea presented in Figure~3 addresses how the \al ic fluctuations from coronal jets might lead to
\al ic-pulse packets on magnetic fields.  Furthermore, our picture provides an explanation for why the
pulse field's angle of inclination to the radial field would increase with radial distance from the Sun,
which tendency has been observed \citep{mozer.et20}.  It is however, not clear to us how the angle
could be increased to an angle of much more than 90$^\circ$, such that the field literally ``switches
back" on itself (e.g., as in extended data Figure~2 of
\citeauthor{kasper.et19}~\citeyear{kasper.et19}). It seems however that only a small percentage of
switchbacks have such rotation angles far beyond 90$^\circ$ \citep{mozer.et20}.  Perhaps a non-linear
and/or turbulent effect, or some additional process in the solar wind, could augment the progression
pictured in Figure~3(c), so that the  field's angle greatly exceeds  90$^\circ$ in some cases.

Our suggested connection between coronal jets and switchbacks is, however, still speculation, and
therefore other ideas cannot be ruled out \citep[e.g.,][]{tenerani.et20}.  Mapping a switchback, perhaps 
a particularly
large one, back along a Parker spiral to a magnetic footpoint on which a jet or series of
jets is observed with the proper timing would provide support for this idea.  In addition, we hope that simulations 
of coronal jets that include the magnetic connections between the solar surface and the heliosphere 
\citep[e.g.][]{lionello.et16,roberts.et18}, with the 
addition of driving the event by a minifilament-field eruption, will be able to test these ideas.

\acknowledgments

We thank an anonymous referee for helpful comments.  This work was supported by funding from the
Heliophysics Division of NASA's Science Mission Directorate through the Heliophysics Guest
Investigators (HGI) Program, and the MSFC \hinode\ Project.

\bibliography{switchbacks_refs}

\begin{thebibliography}{}
\expandafter\ifx\csname natexlab\endcsname\relax\def\natexlab#1{#1}\fi
\providecommand{\url}[1]{\href{#1}{#1}}
\providecommand{\dodoi}[1]{doi:~\href{http://doi.org/#1}{\nolinkurl{#1}}}
\providecommand{\doeprint}[1]{\href{http://ascl.net/#1}{\nolinkurl{http://ascl.net/#1}}}
\providecommand{\doarXiv}[1]{\href{https://arxiv.org/abs/#1}{\nolinkurl{https://arxiv.org/abs/#1}}}

\bibitem[{Alzate \& Morgan(2016)}]{alzate.et16}
Alzate, N., \& Morgan, H. 2016, Astrophysical Journal, 823, 129,
  \dodoi{10.3847/0004-637X/823/2/129}

\bibitem[{Bale {et~al.}(2016)Bale, Goetz, Harvey, Turin, Bonnell, Dudok~de Wit,
  Ergun, MacDowall, Pulupa, Andre, Bolton, Bougeret, Bowen, Burgess, Cattell,
  Chandran, Chaston, Chen, Choi, Connerney, Diaz-Aguado, Donakowski, Drake,
  Farrell, Fergeau, Fermin, Fischer, Fox, Glaser, Goldstein, Gordon, Hanson,
  Harris, Hayes, Hinze, Hollweg, Horbury, Howard, Hoxie, Jannet, Karlsson,
  Kasper, Kellogg, Kien, Klimchuk, Krasnoselskikh, Krucker, Lynch, Maksimovic,
  Malaspina, Marker, Martin, Martinez-Oliveros, McCauley, McComas, McDonald,
  Meyer-Vernet, Moncuquet, Monson, Mozer, Murphy, Odom, Oliverson, Olson,
  Parker, Pankow, Phan, Quataert, Quinn, Ruplin, Salem, Seitz, Sheppard, Siy,
  Stevens, Summers, Szabo, Timofeeva, Vaivads, Velli, Yehle, Werthimer, \&
  Wygant}]{bale.et16}
Bale, S.~D., Goetz, K., Harvey, P.~R., {et~al.} 2016, Space Science Reviews,
  204, 49, \dodoi{10.1007/s11214-016-0244-5}

\bibitem[{Bale {et~al.}(2019)Bale, Badman, Bonnell, Bowen, Burgess, Case,
  Cattell, Chandran, Chaston, Chen, Drake, Dudok~de Witt, Eastwood, Ergun,
  Farrell, Fong, Goetz, Goldstein, Goodrich, Harvey, Horbury, Howes, Kasper,
  Kellogg, Klimcuk, Korreck, Krasnoselskikh, Krucker, Laker, Larson, MacDowall,
  Maksimovic, Malaspina, Martinez-Oliveros, McComas, Meyer-Vernet, Moncuquet,
  Mozer, Phan, Pulupa, Raouafi, Salem, Stansby, Stevens, Szabo, Velli, Woolley,
  \& Wygant}]{bale.et19}
Bale, S.~D., Badman, S.~T., Bonnell, J.~W., {et~al.} 2019, Nature, 576, 237,
  \dodoi{10.1038/s41586-019-1818-7}

\bibitem[{Bemporad {et~al.}(2005)Bemporad, Sterling, Moore, \&
  Poletto}]{bemporad.et05}
Bemporad, A., Sterling, A.~C., Moore, R.~L., \& Poletto, G. 2005, Astrophysical
  Journal, 635L, 189, \dodoi{10.1086/499625}

\bibitem[{Chen {et~al.}(2012)Chen, Zhang, \& Ma}]{chen.et12}
Chen, H., Zhang, J., \& Ma, S. 2012, Research in Astronomy and Astrophysics,
  12, 573, \dodoi{10.1088/1674-4527/12/5/009}

\bibitem[{Cirtain {et~al.}(2007)Cirtain, Golub, Lundquist, van Ballegooijen,
  Savcheva, Shimojo, DeLuca, Tsuneta, Sakao, Reeves, Weber, Kano, Narukage, \&
  Shibasaki}]{cirtain.et07}
Cirtain, J.~W., Golub, L., Lundquist, L., {et~al.} 2007, Science, 318, 1580,
  \dodoi{10.1126/science.1147050}

\bibitem[{Duan {et~al.}(2019)Duan, Shen, Chen, \& Liang}]{duan.et19}
Duan, Y., Shen, Y., Chen, H., \& Liang, H. 2019, Astrophysical Journal, 881,
  132, \dodoi{10.3847/1538-4357/ab32e9}

\bibitem[{Dudok~de Wit {et~al.}(2020)Dudok~de Wit, Krasnoselskikh, Bale,
  Bonnell, Bowen, Chen, Froment, Goetz, Harvey, Jagarlamudi, Larosa, MacDowall,
  Malaspina, Matthaeus, Pulupa, Velli, \& Whittlesey}]{dudok.et20}
Dudok~de Wit, T., Krasnoselskikh, V.~V., Bale, S.~D., {et~al.} 2020,
  Astrophysical Journal Supplement Series, 246, 39,
  \dodoi{10.3847/1538-4365/ab5853}

\bibitem[{Fox {et~al.}(2016)Fox, Velli, Bale, Decker, Driesman, Howard, Kasper,
  Kinnison, Kusterer, Lario, Lockwood, McComas, Raouafi, \& Szabo}]{fox.et16}
Fox, N.~J., Velli, M.~C., Bale, S.~D., {et~al.} 2016, Space Science Reviews,
  204, 7, \dodoi{10.1007/s11214-015-0211-6}

\bibitem[{{Hinode Review Team} {et~al.}(2019){Hinode Review Team}, {Khalid},
  {Patrick}, {Baker}, R., \& {et al.}}]{hinode.et19}
{Hinode Review Team}, {Khalid}, A.-J., {Patrick}, A., {et~al.} 2019,
  Publications of the Astronomical Society of Japan, 71, id.R1,
  \dodoi{10.1093/pasj/psz084}

\bibitem[{Hong {et~al.}(2014)Hong, Jiang, Yang, Bi, Li, Yang, \&
  Yang}]{hong.et14}
Hong, J., Jiang, Y., Yang, J., {et~al.} 2014, Astrophysical Journal, 796, 73,
  \dodoi{10.1088/0004-637X/796/2/73}

\bibitem[{Hong {et~al.}(2013)Hong, Jiang, Yang, Zheng, Bi, Li, Yang, \&
  Yang}]{hong.et13}
---. 2013, Research in Astronomy and Astrophysics, 13, 253,
  \dodoi{10.1088/1674-4527/13/3/001}

\bibitem[{Horbury {et~al.}(2020)Horbury, Woolley, Laker, Matteini, Eastwood,
  Bale, Velli, Chandran, Phan, Raouafi, Goetz, Harvey, Pulupa, Klein, Dudok~de
  Wit, Kasper, Korreck, Case, Stevens, Whittlesey, Larson, MacDowall,
  Malaspina, \& Livi}]{horbury.et20}
Horbury, T.~S., Woolley, T., Laker, R., {et~al.} 2020, Astrophysical Journal
  Supplement Series, 246, 45, \dodoi{10.3847/1538-4365/ab5b15}

\bibitem[{Howard {et~al.}(2008)Howard, Moses, Vourlidas, Newmark, Socker,
  Plunkett, Korendyke, Cook, Hurley, Davila, Thompson, St~Cyr, Mentzell,
  Mehalick, Lemen, Wuelser, Duncan, Tarbell, Wolfson, Moore, Harrison, Waltham,
  Lang, Davis, Eyles, Mapson-Menard, Simnett, Halain, Defise, Mazy, Rochus,
  Mercier, Ravet, Delmotte, Auchere, Delaboudiniere, Bothmer, Deutsch, Wang,
  Rich, Cooper, Stephens, Maahs, Baugh, McMullin, \& Carter}]{howard.et08}
Howard, R.~A., Moses, J.~D., Vourlidas, A., {et~al.} 2008, Space Science
  Reviews, 136, 67, \dodoi{10.1007/s11214-008-9341-4}

\bibitem[{Kahler {et~al.}(1996)Kahler, Crooker, \& Gosling}]{kahler.et96}
Kahler, S.~W., Crooker, N.~U., \& Gosling, J.~T. 1996, Journal of Geophysical
  Research, 101, 24373, \dodoi{10.1029/96JA02232}

\bibitem[{Kamio {et~al.}(2010)Kamio, Curdt, Teriaca, Inhester, \&
  Solanki}]{kamio.et10}
Kamio, S., Curdt, W., Teriaca, L., Inhester, B., \& Solanki, S.~K. 2010,
  Astronomy and Astrophysics, 510, 1, \dodoi{10.1051/0004-6361/200913269}

\bibitem[{Kasper {et~al.}(2016)Kasper, Abiad, Austin, Balat-Pichelin, Bale,
  Belcher, Berg, Bergner, Berthomier, Bookbinder, Brodu, Caldwell, Case,
  Chandran, Cheimets, Cirtain, Cranmer, Curtis, Daigneau, Dalton, DeTomaso,
  Diaz-Aguado, Djordjevic, Donaskowski, Effinger, Florinski, Fox, Freeman,
  Gallagher, Gary, Gauron, Gates, Goldstein, Golub, Gordon, Gurnee, Guth,
  Halekas, Hatch, Heerikuisen, Ho, Hu, Johnson, Jordan, Korreck, Larson,
  Lazarus, Li, Livi, Ludlam, Maksimovic, McFadden, Marchant, Maruca, McComas,
  Messina, Mercer, Park, Peddie, Pogorelov, Reinhart, Richardson, Robinson,
  Rosen, Skoug, Slagle, Steinberg, Stevens, Szabo, Taylor, Tiu, Turin, Velli,
  Webb, Whittlesey, Wright, Wu, \& Zank}]{kasper.et16}
Kasper, J.~C., Abiad, R., Austin, G., {et~al.} 2016, Space Science Reviews,
  204, 131, \dodoi{2016SSRv..204..131K}

\bibitem[{Kasper {et~al.}(2019)Kasper, Bale, Belcher, Berthomier, Case,
  Chandran, Curtis, Gallagher, Gary, Golub, Halekas, Ho, Horbury, Hu, Huang,
  Klein, Korreck, Larson, Livi, Maruca, Lavraud, Louarn, Maksimovic,
  Martinovic, McGinnis, Pogorelov, Richardson, Skoug, Steinberg, Stevens,
  Szabo, Velli, Whittlesey, Wright, Zank, MacDowall, McComas, McNutt, Pulupa,
  Raouafi, \& Schwadron}]{kasper.et19}
Kasper, J.~C., Bale, S.~D., Belcher, J.~W., {et~al.} 2019, Nature, 576, 228,
  \dodoi{10.1038/s41586-019-1813-z}

\bibitem[{Kumar {et~al.}(2018)Kumar, Karpen, Antiochos, Wyper, DeVore, \&
  DeForest}]{kumar.et19}
Kumar, P., Karpen, J.~T., Antiochos, S.~K., {et~al.} 2018, Astrophysical
  Journal, 873, 93, \dodoi{10.3847/1538-4357/ab04af}

\bibitem[{Lionello {et~al.}(2016)Lionello, T{\" o}r{\" o}k, Titov, Leake,
  Miki{\' c}, Linker, \& Linton}]{lionello.et16}
Lionello, R., T{\" o}r{\" o}k, T., Titov, V.~S., {et~al.} 2016, Astrophysical
  Journal, 831L, 2, \dodoi{10.3847/2041-8205/831/1/L2}

\bibitem[{Liu {et~al.}(2019)Liu, Wang, \& Erd{\'e}lyi}]{liu.et19}
Liu, J., Wang, Y., \& Erd{\'e}lyi, R. 2019, Frontiers in Astronomy and Space
  Sciences, 6, 44L, \dodoi{10.3389/fspas.2019.00044}

\bibitem[{McGlasson {et~al.}(2019)McGlasson, Panesar, Sterling, \&
  Moore}]{mcglasson.et19}
McGlasson, R.~A., Panesar, N.~K., Sterling, A.~C., \& Moore, R.~L. 2019, \apj,
  882, 16, \dodoi{10.3847/1538-4357/ab2fe3}

\bibitem[{Miao {et~al.}(2018)Miao, Liu, Li, Shen, Yang, Elmhamdi, Kordi, \&
  Abidin}]{miao.et18}
Miao, Y., Liu, Y., Li, H.~B., {et~al.} 2018, Astrophysical Journal, 869, 39,
  \dodoi{10.3847/1538-4357/aaeac1}

\bibitem[{Moore {et~al.}(2010)Moore, Cirtain, Sterling, \&
  Falconer}]{moore.et10}
Moore, R.~L., Cirtain, J.~W., Sterling, A.~C., \& Falconer, D.~A. 2010,
  Astrophysical Journal, 720, 757, \dodoi{10.1088/0004-637X/720/1/757}

\bibitem[{Moore {et~al.}(2013)Moore, Sterling, Falconer, \& Robe}]{moore.et13}
Moore, R.~L., Sterling, A.~C., Falconer, D.~A., \& Robe, D. 2013, Astrophysical
  Journal, 769, 134, \dodoi{10.1088/0004-637X/769/2/134}

\bibitem[{Moore {et~al.}(2015)Moore, Sterling, \& Falconer}]{moore.et15}
Moore, R.~L., Sterling, R.~L., \& Falconer, D.~A. 2015, Astrophysical Journal,
  806, 11, \dodoi{10.1088/0004-637X/806/1/11}

\bibitem[{Moore {et~al.}(2018)Moore, Sterling, \& Panesar}]{moore.et18}
Moore, R.~L., Sterling, R.~L., \& Panesar, N.~K. 2018, Astrophysical Journal,
  859, 3, \dodoi{10.3847/1538-4357/aabe79}

\bibitem[{Mozer {et~al.}(2020)Mozer, Agapitov, Bale, Bonnell, Case, Chaston,
  Curtis, Dudok~de Wit, Goetz, Goodrich, Harvey, Kasper, Korreck,
  Krasnoselskikh, Larson, Livi, MacDowall, Malaspina, Pulupa, Stevens,
  Whittlesey, \& Wygant}]{mozer.et20}
Mozer, F.~S., Agapitov, O.~V., Bale, S.~D., {et~al.} 2020, Astrophysical
  Journal Supplement Series, 246, 68, \dodoi{10.3847/1538-4365/ab7196}

\bibitem[{Nistic{\`o} {et~al.}(2009)Nistic{\`o}, Bothmer, Patsourakos, \&
  Zimbardo}]{nistico.et09}
Nistic{\`o}, G., Bothmer, V., Patsourakos, S., \& Zimbardo, G. 2009, Solar
  Physics, 259, 87, \dodoi{10.1007/s11207-009-9424-8}

\bibitem[{Nistic{\`o} {et~al.}(2010)Nistic{\`o}, Bothmer, Patsourakos, \&
  Zimbardo}]{nistico.et10}
---. 2010, Annales Geophysicae, 28, 687, \dodoi{10.5194/angeo-28-687-2010}

\bibitem[{Panesar {et~al.}(2016{\natexlab{a}})Panesar, Sterling, \&
  Moore}]{panesar.et16b}
Panesar, N.~K., Sterling, A.~C., \& Moore, R.~L. 2016{\natexlab{a}},
  Astrophysical Journal, 822L, 7, \dodoi{10.3847/2041-8205/822/2/L23}

\bibitem[{Panesar {et~al.}(2017)Panesar, Sterling, \& Moore}]{panesar.et17}
---. 2017, Astrophysical Journal, 844, 131, \dodoi{10.3847/1538-4357/aa7b77}

\bibitem[{Panesar {et~al.}(2018{\natexlab{a}})Panesar, Sterling, \&
  Moore}]{panesar.et18a}
---. 2018{\natexlab{a}}, Astrophysical Journal, 853, 189,
  \dodoi{10.3847/1538-4357/aaa3e9}

\bibitem[{Panesar {et~al.}(2016{\natexlab{b}})Panesar, Sterling, Moore, \&
  Chakrapani}]{panesar.et16a}
Panesar, N.~K., Sterling, A.~C., Moore, R.~L., \& Chakrapani, P.
  2016{\natexlab{b}}, Astrophysical Journal, 832L, 7,
  \dodoi{10.3847/2041-8205/832/1/L7}

\bibitem[{Panesar {et~al.}(2018{\natexlab{b}})Panesar, Sterling, Moore, Tiwari,
  De~Pontieu, \& Norton}]{panesar.et18b}
Panesar, N.~K., Sterling, A.~C., Moore, R.~L., {et~al.} 2018{\natexlab{b}},
  Astrophysical Journal, 868L, 27, \dodoi{10.3847/2041-8213/aaef37}

\bibitem[{Paraschiv {et~al.}(2010)Paraschiv, Lacatus, Badescu, Lupu, Simon,
  Mierla, \& Rusu}]{paraschiv.et10}
Paraschiv, A.~R., Lacatus, D.~A., Badescu, T., {et~al.} 2010, Solar Physics,
  264, 365, \dodoi{10.1007/s11207-010-9584-6}

\bibitem[{Pike \& Mason(1998)}]{pike.et98}
Pike, C.~D., \& Mason, H.~E. 1998, Solar Physics, 182, 333,
  \dodoi{10.1023/A:1005065704108}

\bibitem[{Raouafi \& Stenborg(2014)}]{raouafi.et14}
Raouafi, N.~E., \& Stenborg, G. 2014, Astrophysical Journal, 787, 118,
  \dodoi{10.1088/0004-637X/787/2/118}

\bibitem[{Raouafi {et~al.}(2016)Raouafi, Patsourakos, Pariat, Young, Sterling,
  Savcheva, Shimojo, Moreno-Insertis, DeVore, Archontis, Török, Mason, Curdt,
  Meyer, Dalmasse, \& Matsui}]{raouafi.et16}
Raouafi, N.~E., Patsourakos, S., Pariat, E., {et~al.} 2016, Space Science
  Reviews, 201, 1, \dodoi{10.1007/s11214-016-0260-5}

\bibitem[{Roberts {et~al.}(2018)Roberts, Uritsky, DeVore, \&
  Karpen}]{roberts.et18}
Roberts, M.~A., Uritsky, V.~M., DeVore, C.~R., \& Karpen, J.~T. 2018,
  Astrophysical Journal, 866, 14, \dodoi{10.3847/1538-4357/aadb41}

\bibitem[{Savcheva {et~al.}(2007)Savcheva, Cirtain, Deluca, Lundquist, Golub,
  Weber, Shimojo, Shibasaki, Sakao, Narukage, Tsuneta, \& Kano}]{savcheva.et07}
Savcheva, A., Cirtain, J., Deluca, E.~E., {et~al.} 2007, Publications of the
  Astronomical Society of Japan, 59, 771, \dodoi{10.1093/pasj/59.sp3.S771}

\bibitem[{Shen {et~al.}(2011{\natexlab{a}})Shen, Liu, \& Liu}]{shen.et11b}
Shen, Y., Liu, Y., \& Liu, R. 2011{\natexlab{a}}, Research in Astronomy and
  Astrophysics, 11, 594, \dodoi{10.1088/1674-4527/11/5/009}

\bibitem[{Shen {et~al.}(2012)Shen, Liu, Su, \& Deng}]{shen.et12}
Shen, Y., Liu, Y., Su, J., \& Deng, Y. 2012, Astrophysical Journal, 745, 164,
  \dodoi{10.1088/0004-637X/745/2/164}

\bibitem[{Shen {et~al.}(2011{\natexlab{b}})Shen, Liu, Su, \&
  Ibrahim}]{shen.et11a}
Shen, Y., Liu, Y., Su, J., \& Ibrahim, A. 2011{\natexlab{b}}, Astrophysical
  Journal, 735L, 43, \dodoi{10.1088/2041-8205/735/2/L43}

\bibitem[{Shen {et~al.}(2017)Shen, Liu, Su, Qu, \& Tian}]{shen.et17}
Shen, Y., Liu, Y.~D., Su, J., Qu, Z., \& Tian, Z. 2017, Astrophysical Journal,
  851, 67, \dodoi{10.3847/1538-4357/aa9a48}

\bibitem[{Shen {et~al.}(2019)Shen, Qu, Zhou, Duan, Tang, \& Yuan}]{shen.et19}
Shen, Y., Qu, Z., Zhou, C., {et~al.} 2019, Astrophysical Journal, 855L, 11,
  \dodoi{10.3847/2041-8213/ab4cf3}

\bibitem[{Shibata \& Magara(2011)}]{shibata.et11}
Shibata, K., \& Magara, T. 2011, LRSP, 8, 6

\bibitem[{Shibata \& Uchida(1986)}]{shibata.et86}
Shibata, K., \& Uchida, Y. 1986, Solar Physics, 178, 379

\bibitem[{Shibata {et~al.}(1992)Shibata, Ishido, Acton, Strong, Hirayama,
  Uchida, McAllister, Matsumoto, Tsuneta, Shimizu, Hara, Sakurai, Ichimoto,
  Nishino, \& Ogawara}]{shibata.et92}
Shibata, K., Ishido, Y., Acton, L.~W., {et~al.} 1992, Publications of the
  Astronomical Society of Japan, 44, L173

\bibitem[{Solanki {et~al.}(2019)Solanki, Srivastava, Rao, \&
  Dwivedi}]{solanki.et19}
Solanki, R., Srivastava, A.~K., Rao, Y.~K., \& Dwivedi, B.~N. 2019, Solar
  Physics, 294, 68, \dodoi{10.1007/s11207-019-1453-3}

\bibitem[{Sterling {et~al.}(2015)Sterling, Moore, Falconer, \&
  Adams}]{sterling.et15}
Sterling, A.~C., Moore, R.~L., Falconer, D.~A., \& Adams, M. 2015, Nature, 523,
  437, \dodoi{10.1038/nature14556}

\bibitem[{Sterling {et~al.}(2016)Sterling, Moore, Falconer, Panesar, Akiyama,
  Yashiro, \& Gopalswamy}]{sterling.et16b}
Sterling, A.~C., Moore, R.~L., Falconer, D.~A., {et~al.} 2016, Astrophysical
  Journal, 821, 100, \dodoi{10.3847/0004-637X/821/2/100}

\bibitem[{Sterling {et~al.}(2017)Sterling, Moore, Falconer, Panesar, \&
  Martinez}]{sterling.et17}
Sterling, A.~C., Moore, R.~L., Falconer, D.~A., Panesar, N.~K., \& Martinez, F.
  2017, Astrophysical Journal, 844, 28, \dodoi{10.3847/1538-4357/aa7945}

\bibitem[{Sterling {et~al.}(2018)Sterling, Moore, \& Panesar}]{sterling.et18}
Sterling, A.~C., Moore, R.~L., \& Panesar, N.~K. 2018, Astrophysical Journal,
  864, 68, \dodoi{10.3847/1538-4357/aad550}

\bibitem[{Suess(2007)}]{suess07}
Suess, S. 2007, in Proceedings of The Second Solar Orbiter Workshop, ed.
  K.~Marsch, E.;~Tsinganos \& L.~Marsden, R.;~Conroy (Noordwijk, Netherlands:
  European Space Agency), 641

\bibitem[{Tenerani {et~al.}(2020)Tenerani, Velli, Matteini, R{\'e}ville, Shi,
  Bale, Kasper, Bonnell, Case, de~Wit, Goetz, Harvey, Klein, Korreck, Larson,
  Livi, MacDowall, Malaspina, Pulupa, \& Stevens}]{tenerani.et20}
Tenerani, A., Velli, M., Matteini, L., {et~al.} 2020, Astrophysical Journal
  Supplement Series, 246, 32, \dodoi{10.3847/1538-4365/ab53e1}

\bibitem[{Wang {et~al.}(1998)Wang, Sheeley, Jr., Socker, G., Howard, Brueckner,
  Michels, Moses, Cyr, C., Llebaria, \& Delaboudinière}]{wang.et98}
Wang, Y.-M., Sheeley, N.~R., Jr., {et~al.} 1998, Astrophysical Journal, 508,
  899, \dodoi{10.1086/306450}

\bibitem[{Wyper {et~al.}(2017)Wyper, Antiochos, \& DeVore}]{wyper.et17}
Wyper, P.~F., Antiochos, S.~K., \& DeVore, C.~R. 2017, Nature, 544, 452,
  \dodoi{10.1038/nature22050}

\bibitem[{Wyper {et~al.}(2018)Wyper, DeVore, \& Antiochos}]{wyper.et18a}
Wyper, P.~F., DeVore, C.~R., \& Antiochos, S.~K. 2018, Astrophysical Journal,
  852, 98, \dodoi{10.3847/1538-4357/aa9ffc}

\bibitem[{Yamauchi {et~al.}(2004)Yamauchi, Suess, Steinberg, \&
  Sakurai}]{yamauchi.et04}
Yamauchi, Y., Suess, S.~T., Steinberg, J.~T., \& Sakurai, T. 2004, Journal of
  Geophysical Research: Space Physics, 109, A03104,
  \dodoi{10.1029/2003JA010274}

\bibitem[{Yokoyama \& Shibata(1995)}]{yokoyama.et95}
Yokoyama, T., \& Shibata, K. 1995, Nature, 375, 42, \dodoi{10.1038/375042a0}

\bibitem[{Young \& Muglach(2014{\natexlab{a}})}]{young.et14a}
Young, P.~R., \& Muglach, K. 2014{\natexlab{a}}, Solar Physics, 289, 3313,
  \dodoi{10.1007/s11207-014-0484-z}

\bibitem[{Young \& Muglach(2014{\natexlab{b}})}]{young.et14b}
---. 2014{\natexlab{b}}, Publications of the Astronomical Society of Japan, 66,
  12, \dodoi{10.1093/pasj/psu088}

\bibitem[{Yu {et~al.}(2014)Yu, Jackson, Buffington, Hick, Shimojo, \&
  Sako}]{yu.et14}
Yu, H.-S., Jackson, B.~V., Buffington, A., {et~al.} 2014, Astrophysical
  Journal, 784, 166, \dodoi{10.1088/0004-637X/784/2/166}

\bibitem[{Yu {et~al.}(2016)Yu, Jackson, Yang, Chen, Buffington, \&
  Hick}]{yu.et16}
Yu, H.-S., Jackson, B.~V., Yang, Y.~H., {et~al.} 2016, Journal of Geophysical
  Research: Space Physics, 121, 4985, \dodoi{10.1002/2016JA022503}

\end{thebibliography}

\clearpage


\begin{deluxetable}{cccccclc}
\tabletypesize{\footnotesize}
\tablecaption{Jets in AIA 304~\AA\ Movie \label{tab:table1}}
\tablehead{
\colhead{Event} & \colhead{Prev.~Event\tablenotemark{a}} &\colhead{Start\tablenotemark{b}} & \colhead{End\tablenotemark{c}} & \colhead{Duration [min]\tablenotemark{d}} & 
\colhead{Velocity [\kms]\tablenotemark{e}} & \colhead{Rotations (time period)\tablenotemark{f}} 
& \colhead{COR1 vel [\kms]\tablenotemark{g}}
}
\startdata
J1 & --- & 19:07:20 & --- & --- &  150 & 0.75 (19:14:32--19:19:20) & --- \\ 
J2 & --- & 19:19:56 & --- & --- & 190 & 0.50 (19:24:44--19:29:32) & --- \\
J3 & 5 & 19:30:08 & 19:40:20 & 10 & 255 & 0(?)\tablenotemark{h} & 368$\pm 44$\\
J4 & 6 & 20:08:32  & 20:34:32 & 26 & 255 & 0.75 (20:14:32--20:24:08)& 479$\pm 17$\\
J5 & 7(?)\tablenotemark{i} & 20:37:20  & 21:09:08  & 32 & 170 & 1.5 (20:50:32-- 20:54:44)& 521$\pm 32$ \\
J6 & 8  & 21:17:32  & 21:49:20  & 32 & 615 & 0.25 (21:23:32--21:30:44)& 841 $\pm 10$\\
J7 & -- & 22:57:08  & 23:24:44  & 28 & 270 &  0.5 (23:00:08--23:12:44 & ---\\
J8 & -- & 23:24:44  & 23:43:20  & 19 & 135 &  1.5 (23:25:56--23:28:56)  & ---\\
\hline
Averages & --- &  ---  & ---  & $24.5\pm 8.6$ & $255\pm 155$  & $0.8\pm 0.5$ \\ 
\enddata
\tablenotetext{a}{Corresponding event number in \citet{sterling.et16b}, when determinable.}
\tablenotetext{b}{Time of earliest clear brightening at base of erupting minifilament that makes the jet.}
\tablenotetext{c}{Approximate time that base activity ceases for this event.  Cannot be determined in some cases due to overlap with subsequent activity.}
\tablenotetext{d}{Difference of previous two columns.}
\tablenotetext{e}{Measured in 304~\AA\ images over Fig.~3 FOV, based on movement of bright spire features during fast-rise phase (i.e., following an initial slow start to the minifilament's rise.}
\tablenotetext{f}{Estimated number of 2$\pi$ turns of the spire over time period given in parentheses.}
\tablenotetext{g}{White-light jet velocity in \stereo-B/COR1 coronagraph images, as measured in \citet{sterling.et16b}.}
\tablenotetext{h}{Spinning motion not obvious, but hard to determine with certainty that it does not exist.}
\tablenotetext{i}{The brightening accompanying event~7 in table~1 of \citet{sterling.et16b} 
was from a location west of the Fig.~4 FOV, but our 304~\AA\ jet in the FOV of Fig.~4 likely corresponds 
to the feature listed as jet 7 in the \citet{sterling.et16b} COR1 movie.}

\end{deluxetable}
\clearpage



\begin{figure}
\hspace*{0.0cm}\includegraphics[angle=0,scale=1.05]{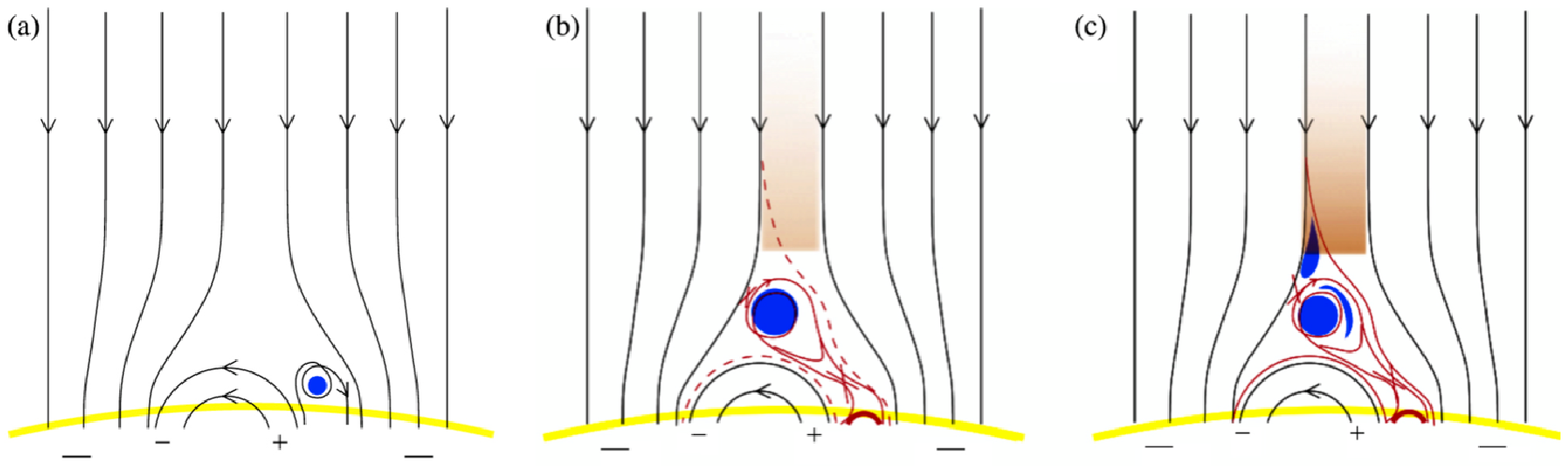}
\caption{
Schematic showing jet generation via a ``minifilament eruption model,'' as proposed in
\citet{sterling.et15}.  This version of the schematic appeared in \citet{sterling.et18}, and 
includes an adjustment due to
\citet{moore.et18}.  (a) Cross-sectional view of a 3D
positive-polarity anemone-type field inside of a majority negative-polarity ambient 
field (which we assume to open into the heliosphere).  One side of the anemone is highly sheared
and contains a minifilament (blue circle).  (b) Here the minifilament is erupting and
undergoing reconnection in two locations: (1) {\it internal} (``tether-cutting'' type)
reconnection (larger red X), with the solid red lines showing the resulting reconnected
fields; the thick red semicircle represents the ``jet bright point'' (JBP\@) at the jet's base;
and (2) 
{\it external} (a.k.a.\ ``interchange'' or ``breakout'' reconnection) occurs at the site of the
smaller red X, with the dashed  lines indicating its two reconnection products.  (c) If the
external reconnection proceeds far enough, then the minifilament material can leak out onto
the open field.  Shaded areas represent heated jet material visible in X-rays and
some \sdo/AIA EUV channels as the jet's spire.  See, e.g., \citet{sterling.et15} or
\citet{moore.et18} for a more detailed description.}
\end{figure}
\clearpage

\begin{figure}
\epsscale{0.7}
\plotone{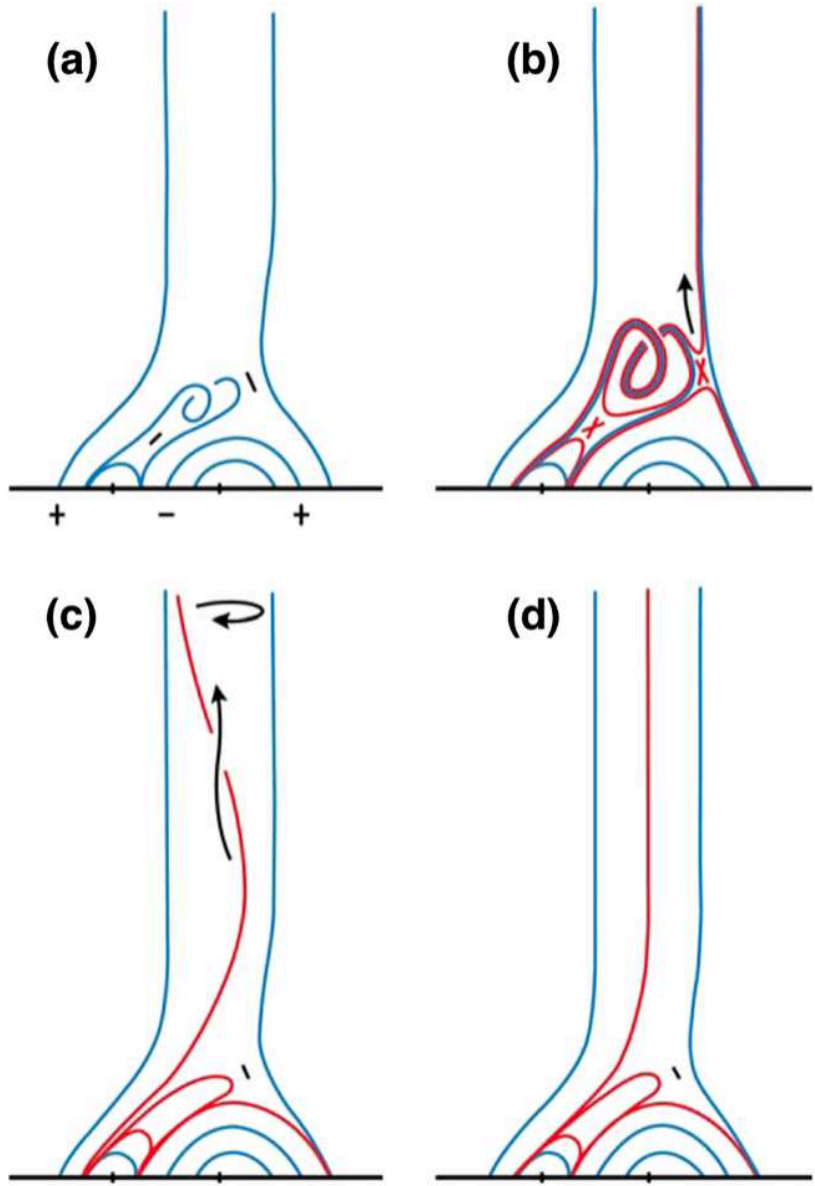}
\caption{Schematic from \citet{moore.et15} of the generation of the magnetic-untwisting wave in an ejective minifilament-eruption
(blowout) jet by the blowout and interchange reconnection of initially closed magnetic field at the base of the jet. At the time
of original publication in \citet{moore.et15}, the full minifilament eruption model (Fig.~1) was still being developed, but 
several critical components of that model are already included here.  Panels~(a) and~(b) show what we now call the minifilament field 
erupting, basically following Fig.~1. In this case however, the schematic emphasizes that  the erupting minifilament field 
contains twist.  That twist is imparted
to the ambient open field via the external reconnection in (b).  This results in a relaxation (untwisting) of
the reconnected twisted ambient coronal field in (c).  Eventually the near-original setup ensues (d), but with the 
original twist in the minifilament field now removed from in and near the jet's base field.  In this representation, the erupting minifilament field has right-handed twist;
this is imparted to the spire field, which then spins in a clockwise direction (viewed from above) to undo the imparted right-handed
twist.}
\end{figure}
\clearpage

\begin{figure}
\hspace*{2.0cm}\includegraphics[angle=0,scale=0.8]{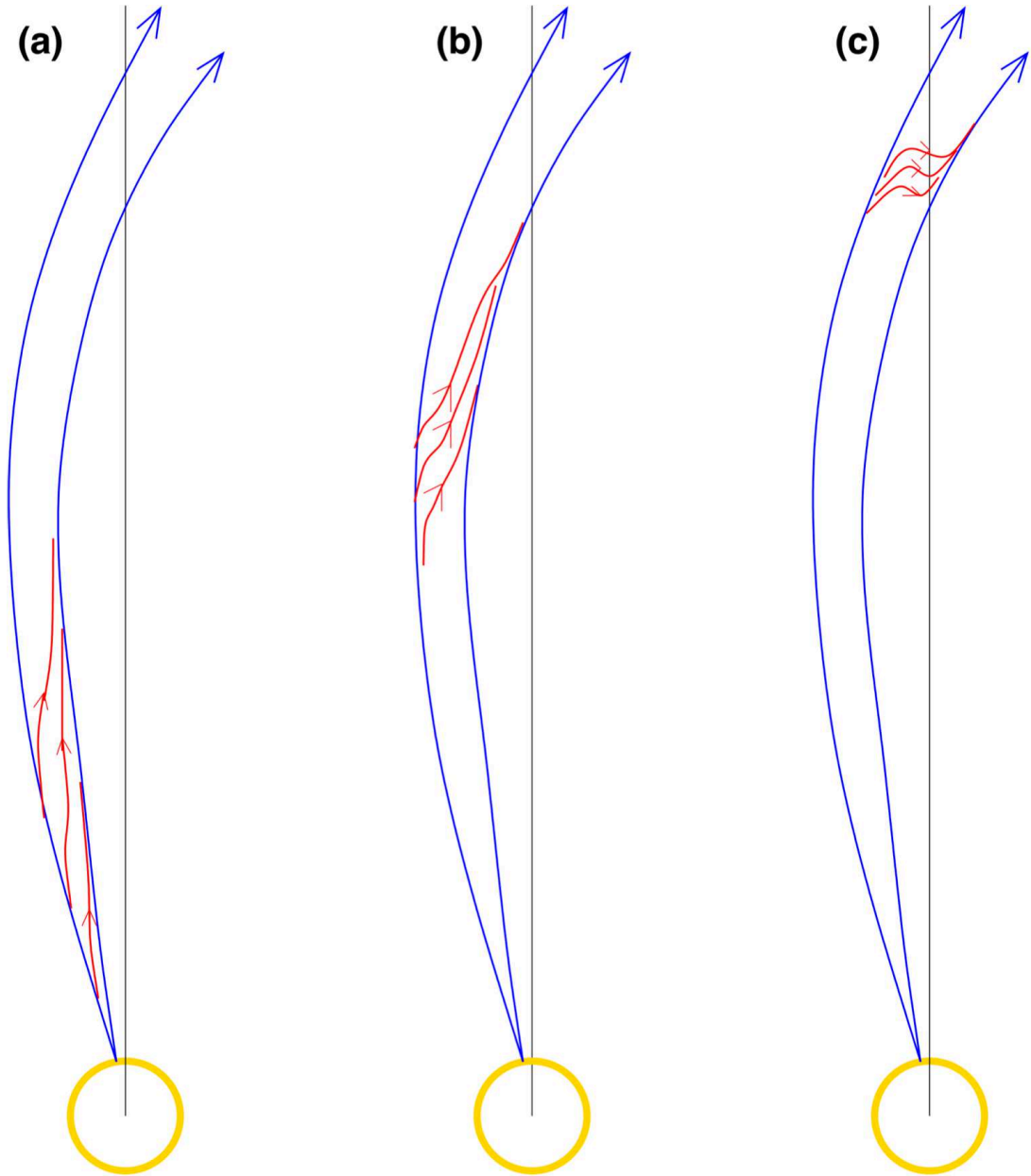}
\caption{Schematic showing a continuation of Fig.~2, where the twist imparted to the ambient coronal field in Fig.~2(c) 
continues to propagate outward.  Here, the yellow circles represent the Sun, and the blue lines represent
heliospheric field lines that are curved with respect to a radial line (black), following a Parker spiral.  In (a),
the wave imparted to the coronal field in Fig.~2(c) becomes the red disturbance, that appears as a
low-pitch twist wave packet moving outward (the radial extent of the twist packet would be comparable to a solar radius, and so 
its extent is exaggerated
by a factor of a few times compared to the Sun  in this schematic representation).  Panel~(b) shows how the pitch of the disturbance is 
expected to increase as it moves further from the Sun, into a regime with lower \al\ speed compared to that in the corona, as described
in the text.  In (c), this pitch-angle steepening of the disturbance continues as it moves even further from the Sun, perhaps
appearing as a ``switchback" by the time it encounters PSP\@.}
\end{figure}
\clearpage

\begin{figure}
\hspace*{0.0cm}\includegraphics[angle=0,scale=1.0]{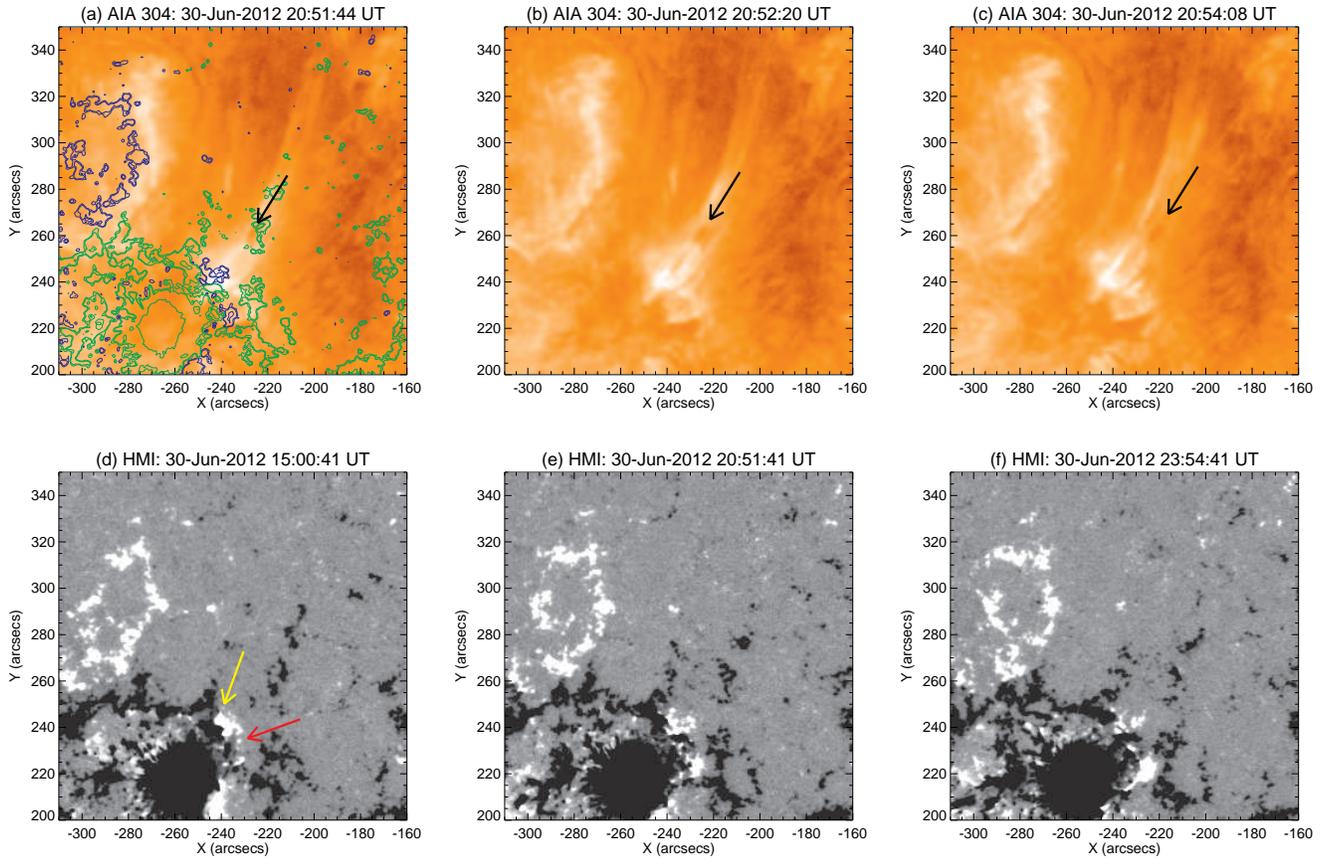}
\caption{Coronal jets from NOAA AR~11513 \citep[studied in detail in][]{sterling.et16b}.  
Panels~(a---c) show \sdo/AIA 304~\AA\ subframes, showing jet J5 of table~1. Arrows show absorbing erupting-filament
material undergoing spinning motion in the successive frames.  Panel~(a) is overlaid with the magnetogram of (e), where 
blue and green contours respectively outline positive and negative polarities.  Panels~(d---f) show \sdo/HMI magnetograms 
of the region, with white and black respectively representing positive and negative polarities.  Arrows in (d) show two
neutral lines that are the source locations of the jets in table~1; the positive-polarity patch between the arrows decreases
with time due to flux cancelation.  According to the model in Fig.~1, this flux cancelation builds the minifilament flux ropes that
erupt to drive the jets, as in (a---c).  Accompanying videos vid4abc and vid4def respectively show time evolution of the 
304~\AA\ images and HMI magnetograms.}
\end{figure}
\clearpage

\begin{figure}
\hspace*{3.0cm}\includegraphics[angle=270,scale=1.0]{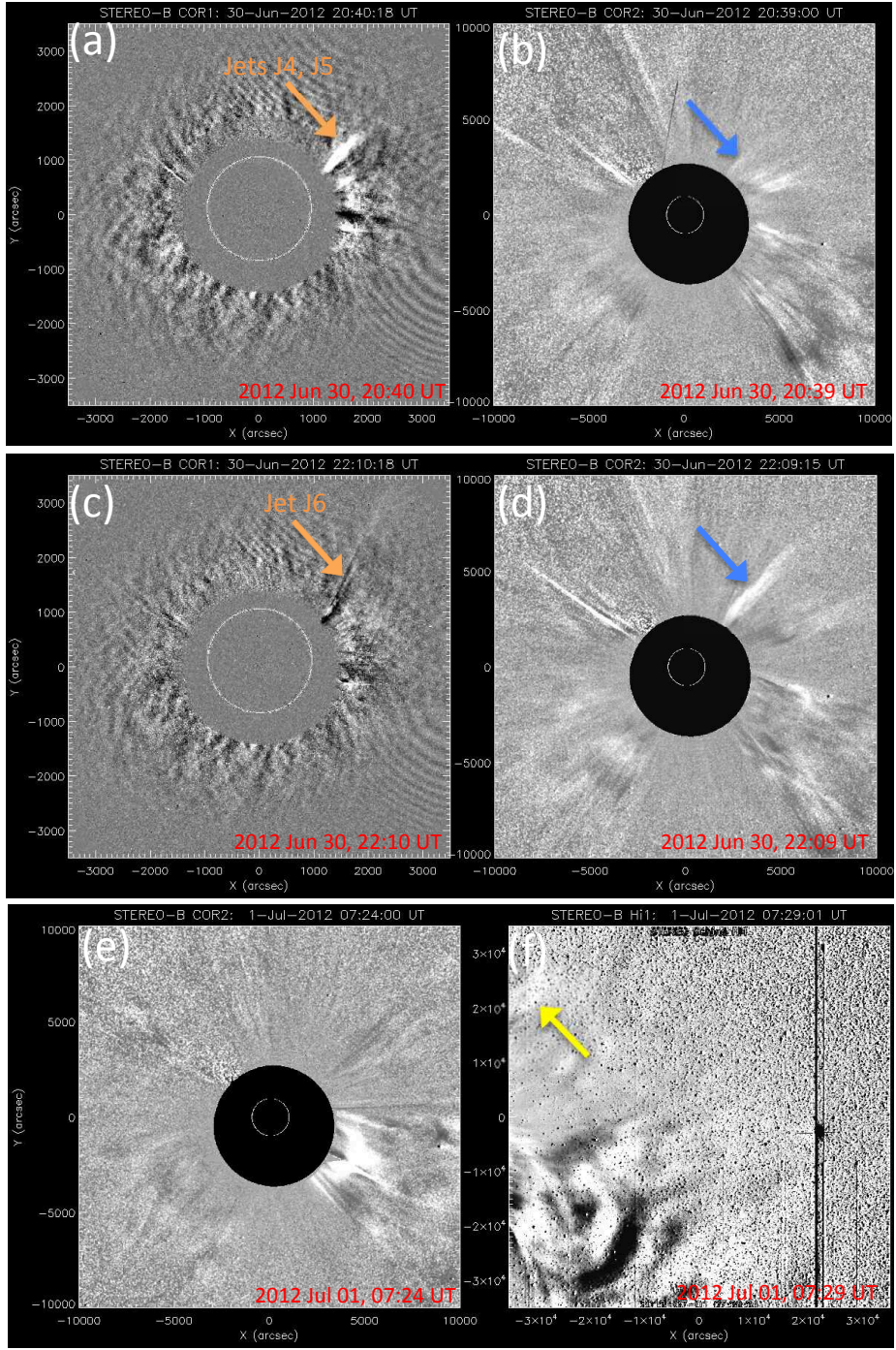}
\caption{Outer-coronal and inner-heliospheric manifestations of coronal jets from AR~11513.  These are coronagraph images
from \stereo-B COR1 (a and c) and COR2 (b, d, e), and a \stereo-B Hi1 heliospheric imager image (f). Horizontal pairs
of images (a)-(b), (c)-(d), and (e)-(f), are respectively at approximately the same times. 
\citet{sterling.et16b} identified the white-light jet in (a) as being due to coronal jets J4 and J5 of table~1 
\citep[jets 6 and 7 of][]{sterling.et16b}, and the white-light in (c) as due to coronal jet J6 of table~1
\citep[jet 8 of][]{sterling.et16b}.  Panels~(b) and (d) show that these white-light jets remain intact (blue arrows) in the COR2
FOV (2.5---15 $R_{\odot}$), and (f) shows that the white-light jet in (d) 
persists into the Hi1 FOV (15---84 $R_{\odot}$ (yellow arrow).  (That jet left the COR2 movie's FOV at about
2:09~UT, and hence is no longer visible in (e).)
Thus it is plausible that the consequences of coronal jets depicted in Fig.~4 can reach PSP locations, and be detected as switchbacks.  Accompanying video vid5abcd shows the time evolution of the 
COR1 and COR2 images, at the 15-min cadence of the available COR2 images (see \citeauthor{sterling.et16b}~\citeyear{sterling.et16b}
for higher-cadence COR1 movie); and accompanying video vid5ef shows the time evolution of the COR2 and Hi1 images, at
the 40-min cadence of the available Hi1 images.}
\end{figure}
\clearpage

\end{document}